\documentclass[twocolumn]{myletters}
\pdfoutput=1
\affiliation{Department of Pure and Applied Mathematics, Waseda University}{3-4-1, Okubo, Shinjuku-ku, Tokyo, 169-8555, Japan}
\authorinfo{Soujun Kitagawa}{}{}
\authorinfo{Daisuke Takahashi}{}{}
\title{Lattice equations and their solutions with complexity of polynomial class}
\abstract{
We discuss initial value problems for time evolution equations in one dimensional space which are expressed by the lattice operators and propose some new equations to which complexity of solutions is of polynomial class. Novel type of expressions using shift operators and binary trees are applied for the derivation of solution.
}
\usepackage{tikz}
\usetikzlibrary{positioning,calc}

\newcommand{\mymax}[2]{
    \draw [line width = 1] (#1,#2) -- ($(#1,#2) + (1,0)$);
    \draw [line width = 1] (#1,#2) -- ($(#1,#2) + (0,-1)$);
    \filldraw [fill = white] (#1,#2) circle [radius = 0.3] node{$\vee$};
}
\newcommand{\mymin}[2]{
    \draw [line width = 1] (#1,#2) -- ($(#1,#2) + (1,0)$);
    \draw [line width = 1] (#1,#2) -- ($(#1,#2) + (0,-1)$);
    \filldraw [fill = white] (#1,#2) circle [radius = 0.3] node {$\wedge$};
}
\newcommand{\myleaf}[3]{
    \fill [white] (#1,#2) circle [radius = 0.3]; \node at (#1,#2) {$#3$};
}
\newcommand{\myhline}[2]{
    \draw [line width = 1] (#1,#2) -- ($(#1,#2) + (1,0)$);
}

\newcommand{\myhdots}[2]{
    \myhline{#1}{#2}
    \fill [white] (#1,#2) circle [radius = 0.4];
    \fill (#1,#2) circle [radius = 0.05];
    \fill ($(0.2,0) + (#1,#2)$) circle [radius = 0.05];
    \fill ($(-0.2,0) + (#1,#2)$) circle [radius = 0.05];
}

\newcommand{\mycon}[2]{
    \fill (#1, #2) circle [radius = 0.1];
}

\begin{document}
\maketitle
\section{Introduction}  \label{sec:intro}
In this letter, we discuss the initial value problem in the form;
\begin{equation}
    u_{i}^{n + 1} = f(u_{i + j_1}^n, \dots, u_{i + j_m}^n), \label{timestep}
\end{equation}
where $f$ is a function of $m$ arguments defined by lattice operators and $u_i^n$ is a state value $u$ dependent on an integer site number $i$ and an integer time $n$. Let us call this type of equations `lattice equation'.  Assume that an initial time is $n = 0$ and we consider a solution $u_i^n \, (n > 0)$ determined by \eqref{timestep} and initial data $\{u_i^0\}$.

General solution can be obtained by applying \eqref{timestep} recursively.
Among equations in the form of \eqref{timestep}, the equations of which complexity of general solution is of polynomial order are reported in the reference~\cite{ikegami}.
We discuss the same class of equations and call it `polynomial class' in this paper.

Complicated procedure on reduction using formulas of lattice operators is necessary to show the complexity of solutions. The first aim of this
paper is to introduce convenient notations on lattice operators and
express them in a form of binary tree. The second aim is to propose
novel equations of polynomial class and show their mechanism on reducing the
complexity using the above notation and expression.
\section{Lattice}
Lattice is an algebraic system with two binary operations, $\vee$ and $\wedge$~\cite{lattice}.
Axioms of lattices are shown below.
\begin{align}
    A \vee A &= A,  \label{idempotent1} \\
    A \wedge A &= A,\label{idempotent2}\\
    A \vee B &= B \vee A,  \label{comlaw1} \\
    A \wedge B &= B \wedge A, \label{comlaw2} \\
    A \vee (B \vee C) &= (A \vee B) \vee C, \\
    A \wedge (B \wedge C) &= (A \wedge B) \wedge C, \\
    A \vee (A \wedge B) &= A, \label{ablaw1}\\
    A \wedge (A \vee B) &= A. \label{ablaw2}
\end{align}
Equations \eqref{ablaw1} and \eqref{ablaw2} are called absorption-law.
Since commutative and associative laws hold, we can use notations $A \vee B \vee C$ and $A \wedge B \wedge C$ denoting $A \vee (B \vee C)$ and $A \wedge (B \wedge C)$ respectively.

Distributive lattice is a lattice such that $\vee$ and $\wedge$ satisfy the following 
axioms in addition to the above.
\begin{align}
    A \vee (B \wedge C) &= (A \vee B) \wedge (A \vee C), \label{dislaw1} \\
    A \wedge (B \vee C) &= (A \wedge B) \vee (A \wedge C). \label{dislaw2}
\end{align}
These are called distribution-law.

Conjugation is an unary operation $A \mapsto \overline{A}$ which satisfies the following axioms.
\begin{align}
    \overline{\overline{A}} &= A, \\
    \overline{A \vee B} &= \overline{A} \wedge \overline{B}, \\
    \overline{A \wedge B} &= \overline{A} \vee \overline{B}.  \label{demorgan}
\end{align}

Since all axioms are symmetric about $\vee$ and $\wedge$,
if an identity including $\vee$ or $\wedge$ holds, then the identity obtained by
replacing $\vee$ and $\wedge$ also holds. For example,
\begin{equation*}
    A \wedge (B \vee (A \wedge \overline{C})) = A \wedge (B \vee \overline{C})
\end{equation*}
is true, then,
\begin{equation*}
    A \vee (B \wedge (A \vee \overline{C})) = A \vee (B \wedge \overline{C})
\end{equation*}
is obtained by exchanging $\vee$ and $\wedge$.

\section{Operator expression}
We discuss initial value problems described by \eqref{timestep} where $u_i^n$ is a state variable of lattice element and $f$ is defined by a combination of lattice operations. The solution $u_i^n$ is obtained by applying \eqref{timestep} recursively to the initial data ${u_i^0}$. We introduce the following special notations to simplify the solution of this recursive applications.

\begin{itemize}
    \item Assume $\varphi_i$ is a combination of operators ($\vee$, $\wedge$, $\overline{\vbox to 5pt{\hbox to 6pt{}}}$) and state variables on one-dimensional space sites with site number $i$.  For example, $\varphi_i=(\overline{u_i}\vee u_{i+3})\wedge u_{i-1}$.
    \item Integer means a shift of site number of all state variables including in $\varphi_i$ by its value. For example, $a \cdot\varphi_i = \varphi_{i + a}$ ($a \in \mathbf{Z}$), $3 \cdot \varphi_i = \varphi_{i + 3}$ and $-2\cdot(u_{i+1}\wedge\overline{u_{i+2}})=u_{i-1}\wedge\overline{u_i}$. We assume that the lower case letter (for example, $a$, $b$, $\dots$) means this type of integer shift.
    \item We use the following operator notations ($a$, $b\in\mathbf{Z}$), 
    \begin{align}
        \overline{a} \cdot \varphi_i &= \overline{\varphi_{i + a}}, \\
        (a \vee b) \cdot \varphi_i &= \varphi_{i + a} \vee \varphi_{i + b}, \\
        (a \wedge b) \cdot \varphi_i &= \varphi_{i + a} \wedge \varphi_{i + b}.
    \end{align}
    Combinations of the above operators can be applied in the same manner. For example, $(a \wedge (\overline{b} \vee c)) \cdot \varphi_i = \varphi_{i + a} \wedge (\overline{\varphi_{i + b}} \vee \varphi_{i + c})$.  Then, any formula obtained from \eqref{idempotent1} to \eqref{demorgan} by replacing elements of lattice by shift operators also holds.  For example, $a\vee(a\wedge b)=a$ and $\overline{a \vee b}=\overline{a} \wedge \overline{b}$.
    \item Combination of operators is denoted by an uppercase letter.
    For example, $F = \overline{2} \vee (-1 \wedge 0)$ and $G=-1 \vee 1$.
    \item Composition of operators is denoted by $\circ$ and defined by $(G\circ F)\cdot\varphi_i=G\cdot(F\cdot \varphi_i)$.  We can obtain the operator expression by executing the composition regardless to $\varphi_i$.  For example, if $F = \overline{2} \vee (-1 \wedge 0)$ and $G = -1 \vee 1$,
\begin{equation*}
  (G\circ F)\cdot\varphi_i=G\cdot(F\cdot\varphi_i)=G\cdot\psi_i=\psi_{i-1}\vee\psi_{i+1},
\end{equation*}
where $\psi_i=F\cdot\varphi_i=\overline{\varphi_{i+2}}\vee(\varphi_{i-1}\wedge\varphi_i)$.  Then we can easily show
    \begin{align*}
        G \circ F =& (-1 \vee 1) \circ (\overline{2} \vee (-1 \wedge 0)) \\
        =& (\overline{(2 - 1)} \vee ((-1 - 1) \wedge (0 - 1))) \\
        &\vee (\overline{(2 + 1)} \vee ((-1 + 1) \wedge (0 + 1))) \\
        =& (\overline{1} \vee (-2 \wedge -1)) \vee (\overline{3} \vee (0 \wedge 1)).
    \end{align*}
\end{itemize}

Composition is associative $(H \circ G) \circ F = H \circ (G \circ F)$. Though shift operator is commutative with any operator as $a \circ F = F \circ a$, operators are not commutative in general, that is, $G \circ F \neq F \circ G$.
Left-distribution $(G_1 \vee G_2) \circ F = (G_1 \circ F) \vee (G_2 \circ F)$ holds. However, 
right-distribution does not hold in general.
$F^n$ is defined recursively by
\begin{equation*}
    F^n =
    \begin{cases}
        0 & (n = 0)\\
        F \circ F^{n - 1} & (n \ge 1)    
    \end{cases},
\end{equation*}
where 0 means an identity operator.
\section{Polynomial class and binary tree}
As mentioned in the section \ref{sec:intro}, we discuss the polynomial class of solutions to lattice equations.  The right-hand side of \eqref{timestep} can be described by an appropriate combination of operators $F$ acting on $u_i^n$, that is,
\begin{equation*}
    f(u_{i + j_1}^n, \dots, u_{i + j_m}^n) = F \cdot u_{i}^n.
\end{equation*}
Note that $F$ does not include any operation on time evolution. Then the time evolution \eqref{timestep} is written by $u_{i}^{n + 1} = F \cdot u_i^n$. The solution of initial value problem is formally given by the initial data as $u_i^n = F^n \cdot u_i^0$. We define the complexity of solution by the number of terms included in the solution.  Thus, the complexity of solution $u_i^n$ is directly related to that of $F^n$.  If the growth of the number is exponential with respect to $n$, we classify the solution into `exponential class'.  If polynomial, `polynomial class'.  The polynomial class is $P_k$ if the growth is $O(n^k)$ at most after the reference \cite{ikegami}.

The above formal solution to \eqref{timestep} contains $m^n$ terms of $\{u_i^0\}$.  However, we can reduce the number of terms using by formulas \eqref{ablaw1}, \eqref{ablaw2}, \eqref{dislaw1} and \eqref{dislaw2}.  For example, consider $F = \overline{-1} \vee 0$. The formal solution contains $2^n$ terms in $F^n$, but $F^2$ can be arranged as following.
\begin{align*}
    F^2 &= (\overline{-1} \circ (\overline{-1} \vee 0)) \vee (0 \circ (\overline{-1} \vee 0)) \\
    &= (-2 \wedge \overline{-1}) \vee (\overline{-1} \vee 0) \\
    &= \overline{-1} \vee 0.
\end{align*}
The rightmost hand is derived from absorption law \eqref{ablaw1}.
Therefore $F^n = \overline{-1} \vee 0$ for any integer $n \ge 1$.

Ikegami et al. reported lattice equations of polynomial class which are generated by simple combination of lattice operators and relate them to elementary cellular automata (ECA) as the special case of state variables\cite{ikegami,wolfram}.

If we judge the class of lattice equations, it is important how the reduction of terms can be applied for the given solution.  However, it is difficult to apply formulas for the complicated expressions constructed from enormous terms and operators.  We introduce a novel expression using binary trees to solve this difficulty.  Since $\vee$ and $\wedge$ are binary operation, any combination of operators can be represented by binary trees.  For example, $F=X \vee (Y \wedge Z)$ is represented by
$
\begin{tikzpicture}[baseline = -1em, x = 2em, y = 2em] 
    \mymax{0}{0} \mymin{1}{0} \myleaf{2}{0}{Z}
    \myleaf{0}{-1}{X} \myleaf{1}{-1}{Y}
\end{tikzpicture}
$ where $X$, $Y$ and $Z$ are arbitrary operators. We place a dot ($\bullet$) on a branch to indicate the conjugation. For example, $\overline{X} \vee (Y \wedge Z)$ can be expressed by 
$
\begin{tikzpicture}[baseline = -1em, x = 2em, y = 2em]
    \mymax{0}{0} \mymin{1}{0} \mycon{0}{-0.5} \myleaf{0}{-1}{X} \myleaf{1}{-1}{Y} \myleaf{2}{0}{Z}
\end{tikzpicture}
$ and
$X \vee \overline{(Y \wedge Z)}$ by
$
\begin{tikzpicture}[baseline = -1em, x = 2em, y = 2em]
    \mymax{0}{0} \mymin{1}{0} \mycon{0.5}{0} \myleaf{0}{-1}{X} \myleaf{1}{-1}{Y} \myleaf{2}{0}{Z}
\end{tikzpicture}
$.

Composition $G \circ F$ is executed replacing any leaf $a$ of $G$ by $a\circ F$ where the composition $a\circ F$ can be executed by replacing any leaf $b$ of $F$ by $a+b$.  The following example shows a composition $F^2=F\circ F$ where $F={0 \wedge 1}$.
\begin{equation*}
        \left(
        \begin{tikzpicture}[baseline = -0.7em, x = 1.4em, y = 1.4em]
            \mymin{0}{0} \myleaf{0}{-1}{0} \myleaf{1}{0}{1}
        \end{tikzpicture}
        \right)^2 =
        \begin{tikzpicture}[baseline = -0.7em, x = 1.4em, y = 1.4em]
            \mymin{0}{0} \mymin{0}{-1} \myhline{1}{0} \myhline{1.5}{0} \mymin{2.5}{0}
            \myleaf{0}{-2}{0} \myleaf{1}{-1}{1} \myleaf{2.5}{-1.1}{(1 + 0)} \myleaf{4.6}{0}{(1 + 1)}
        \end{tikzpicture} \\ =
        \begin{tikzpicture}[baseline = -0.7em, x = 1.4em, y = 1.4em]
            \mymin{0}{0} \mymin{0}{-1} \myhline{1}{0} \myhline{1.5}{0} \mymin{2}{0}
            \myleaf{0}{-2}{0} \myleaf{1}{-1.1}{1} \myleaf{2}{-1.1}{1} \myleaf{3}{0}{2}
        \end{tikzpicture}.
\end{equation*}
\section{Reduction mechanism and polynomial class}
Reduction of number of terms can be realized by lattice formulas, that is, idempotent laws (\eqref{idempotent1}, \eqref{idempotent2}), absorption laws (\eqref{ablaw1}, \eqref{ablaw2}) and distribution laws (\eqref{dislaw1}, \eqref{dislaw2}).  Important reduction schemes in binary trees are obtained by using these formulas as follows.
    \begin{align}        
        \begin{tikzpicture}[baseline = -1em, x = 2em, y = 2em]
            \myhdots{0}{0} \mymax{1}{0} \myleaf{1}{-1}{X} \myhdots{2}{0} \mymax{3}{0} \mymin{4}{0}
            \myleaf{3}{-1}{Y} \myleaf{4}{-1}{X} \myleaf{5}{0}{Z}
            \draw[line width = 1] (2.5,-1.5) -- (5.3,-1.5) -- (5.3, 0.5) -- (2.5, 0.5) -- (2.5,-1.5);
        \end{tikzpicture} &=
        \begin{tikzpicture}[baseline = -1em, x = 2em, y = 2em]
            \myhdots{0}{0} \mymax{1}{0} \myleaf{1}{-1}{X} \myhdots{2}{0} \myleaf{3}{0}{Y}
            \draw[line width = 1] (2.5,-0.5) -- (3.5,-0.5) -- (3.5, 0.5) -- (2.5, 0.5) -- (2.5,-0.5);
        \end{tikzpicture}, \\
        \begin{tikzpicture}[baseline = -1em, x = 2em, y = 2em]
            \myhdots{0}{0} \mymin{1}{0} \myleaf{1}{-1}{X} \myhdots{2}{0} \mymin{3}{0} \mymax{4}{0}
            \myleaf{3}{-1}{Y} \myleaf{4}{-1}{X} \myleaf{5}{0}{Z}
            \draw[line width = 1] (2.5,-1.5) -- (5.3,-1.5) -- (5.3, 0.5) -- (2.5, 0.5) -- (2.5,-1.5);
        \end{tikzpicture} &=
        \begin{tikzpicture}[baseline = -1em, x = 2em, y = 2em]
            \myhdots{0}{0} \mymin{1}{0} \myleaf{1}{-1}{X} \myhdots{2}{0} \myleaf{3}{0}{Y}
            \draw[line width = 1] (2.5,-0.5) -- (3.5,-0.5) -- (3.5, 0.5) -- (2.5, 0.5) -- (2.5,-0.5);
        \end{tikzpicture}. \label{genab}
    \end{align}
Note that the boxed parts in left-hand side can be reduced to those in right-hand side.  An example of reduction by the above schemes is shown as follows. The operator $(\overline{a} \wedge b)^2$ is expanded as follows.
\begin{equation*}
\left(
    \begin{tikzpicture}[baseline = -1em, x = 2em, y = 2em]
        \mymin{0}{0} \mycon{0}{-0.5} \myleaf{0}{-1}{a} \myleaf{1}{0}{b}
    \end{tikzpicture}
\right)^2 = 
\begin{tikzpicture}[baseline = -2em, x = 2em, y = 2em]
    \mymin{0}{0} \mymax{0}{-1} \myleaf{0}{-2}{a \circ a} \mycon{0.65}{-1} \myhline{1}{0} \myhline{1.8}{0} \myleaf{1.6}{-1}{a\circ b} \mymin{3}{0} \mycon{3}{-0.5} \myleaf{3}{-1}{a \circ b} \myleaf{4.6}{0}{b \circ b}
\end{tikzpicture}.
\end{equation*}
The terms of the right-hand side are rearranged by commutative law \eqref{comlaw2} as
\begin{equation*}
    \begin{tikzpicture}[baseline = -1em, x = 2em, y = 2em]
        \mymin{0}{0} \mycon{0}{-0.5} \myleaf{0}{-1}{a \circ b} \myhline{1}{0} \myhline{1.3}{0} \mymin{1.3}{0} \myleaf{1.3}{-1}{b \circ b} \mymax{2.6}{0} \myleaf{2.6}{-1}{a \circ b} \mycon{2.6}{-0.5} \myleaf{4.2}{0}{a \circ a}
        \draw[line width = 1] (0.7,0.5) -- (5,0.5) -- (5,-1.5) -- (0.7,-1.5) -- (0.7,0.5);
    \end{tikzpicture}.
\end{equation*}
If the scheme \eqref{genab} applied to the boxed part, the reduced expression,
\begin{equation*}
\left(
    \begin{tikzpicture}[baseline = -1em, x = 2em, y = 2em]
        \mymin{0}{0} \mycon{0}{-0.5} \myleaf{0}{-1}{a} \myleaf{1}{0}{b}
    \end{tikzpicture}
\right)^2 =
\begin{tikzpicture}[baseline = -1em, x = 2em, y = 2em]
    \mymin{0}{0} \mycon{0}{-0.5} \myleaf{0}{-1}{a \circ b} \myleaf{1.6}{0}{b \circ b}
\end{tikzpicture},
\end{equation*}
can be derived.

Below we give the list of equations of polynomial class and their reduced solutions.  We show the evolution operator $F$ and its reduced expression $F^n$ for $u_i^{n+1}=F\cdot u_i^n$ in the list.  Note that the composition operator $\circ$ is abbreviated in the list.
\subsection{Class $P_0$} \label{sec:P0}
\begin{itemize}
\item[1.]
$F=a$, $F^n=a^n$.
\item[2.]
$F=
        \begin{tikzpicture}[baseline = -1em, x = 2em, y = 2em]
            \mymin{0}{0} \mycon{0.5}{0} \myleaf{0}{-1}{a} \myleaf{1}{0}{b}
        \end{tikzpicture}$,
$F^n = a^{n - 1}
    \left(
        \begin{tikzpicture}[baseline = -1em, x = 2em, y = 2em]
            \mymin{0}{0} \mycon{0.5}{0} \myleaf{0}{-1}{a} \myleaf{1}{0}{b}
        \end{tikzpicture}
    \right)$.
\item[3.]
$F=
\begin{tikzpicture}[baseline = -1em, x = 2em, y = 2em]
            \mymin{0}{0} \mycon{0}{-0.5} \mycon{0.5}{0} \myleaf{0}{-1}{a} \myleaf{1}{0}{b}
\end{tikzpicture}$,
$F^{2n-1}=(a b)^{n - 1}
    \left(
        \begin{tikzpicture}[baseline = -1em, x = 2em, y = 2em]
            \mymin{0}{0} \mycon{0}{-0.5} \mycon{0.5}{0} \myleaf{0}{-1}{a} \myleaf{1}{0}{b}
        \end{tikzpicture}
    \right)$,\\
$F^{2n}=(a b)^{n - 1}
    \left(
        \begin{tikzpicture}[baseline = -1em, x = 2em, y = 2em]
            \mymax{0}{0} \mymin{1}{0} \myleaf{0}{-1}{a b} \myleaf{1}{-1}{a^2} \myleaf{2}{0}{b^2}
        \end{tikzpicture}
    \right)$.
\item[4.]
$F=\begin{tikzpicture}[baseline = -1em, x = 2em, y = 2em]
            \mymin{0}{0} \mymin{1}{0} \mycon{1}{-0.5} \mycon{1.5}{0}
            \myleaf{0}{-1}{a} \myleaf{1}{-1}{b} \myleaf{2}{0}{c}
        \end{tikzpicture}$,
$F^n=a^{n - 1}
    \left(
        \begin{tikzpicture}[baseline = -1em, x = 2em, y = 2em]
            \mymin{0}{0} \mymin{1}{0} \mycon{1}{-0.5} \mycon{1.5}{0}
            \myleaf{0}{-1}{a} \myleaf{1}{-1}{b} \myleaf{2}{0}{c}
        \end{tikzpicture}
    \right)$.
\item[5.]
$F=\begin{tikzpicture}[baseline = -1em, x = 2em, y = 2em]
            \mymin{0}{0} \mymax{1}{0} \mycon{1}{-0.5} \mycon{1.5}{0}
            \myleaf{0}{-1}{a} \myleaf{1}{-1}{b} \myleaf{2}{0}{c}
        \end{tikzpicture}$,
$F^n=a^{n - 1}
    \left(
        \begin{tikzpicture}[baseline = -1em, x = 2em, y = 2em]
            \mymin{0}{0} \mymax{1}{0} \mycon{1}{-0.5} \mycon{1.5}{0}
            \myleaf{0}{-1}{a} \myleaf{1}{-1}{b} \myleaf{2}{0}{c}
        \end{tikzpicture}
    \right)$.
\end{itemize}
\subsection{Class $P_1$} \label{sec:P1}
\begin{itemize}
\item[1.]
$F=\begin{tikzpicture}[baseline = -1em, x = 2em, y = 2em]
            \mymin{0}{0} \myleaf{1}{0}{b} \myleaf{0}{-1}{a}
        \end{tikzpicture}$,
$F^n=\begin{tikzpicture}[baseline = -1em , x = 2em, y = 2em]
            \mymin{0}{0} \myhline{1}{0} \mymin{1.2}{0} \myhline{2}{0} \myhdots{2.2}{0} \mymin{3.2}{0} \myleaf{4.2}{0}{b^n}
            \myleaf{0}{-1}{a^n} \myleaf{1.2}{-1}{a^{n-1} b} \myleaf{3.2}{-1}{a b^{n-1}}
        \end{tikzpicture}$.
\item[2.]
$F=\begin{tikzpicture}[baseline = -1em, x = 2em, y = 2em]
            \mymin{0}{0} \mycon{0}{-0.5} \myleaf{0}{-1}{a} \mymin{1}{0} \myleaf{1}{-1}{b} \myleaf{2}{0}{c}
        \end{tikzpicture}$,\\
$F^n=\left(
        \begin{tikzpicture}[baseline = -1em, x = 2em, y = 2em]
            \mymin{0}{0} \myleaf{1}{0}{c} \myleaf{0}{-1}{b}
        \end{tikzpicture}
    \right)^{n - 1}
    \left(
        \begin{tikzpicture}[baseline = -1em, x = 2em, y = 2em]
            \mymin{0}{0} \mycon{0}{-0.5} \myleaf{0}{-1}{a} \mymin{1}{0} \myleaf{1}{-1}{b} \myleaf{2}{0}{c}
        \end{tikzpicture}
    \right)$.
\item[3.]
$F=\begin{tikzpicture}[baseline = -1em, x = 2em, y = 2em]
            \mymin{0}{0} \mymax{1}{0} \myleaf{2}{0}{c} \myleaf{0}{-1}{a} \mycon{1}{-0.5} \myleaf{1}{-1}{b}
        \end{tikzpicture}$, \\$F^n=$
$\begin{tikzpicture}[baseline = -1em, x = 2em, y = 2em]
            \mymin{0}{0} \myleaf{0}{-1}{a^n} \myhline{1}{0} \mymax{1}{0} \mycon{1}{-0.5} \myleaf{1}{-1}{a^{n - 1} b}
            \myhline{1.7}{0} \mymin{2.3}{0} \myleaf{2.3}{-1}{a^{n - 1} c} \myhline{3.3}{0}
            \mymax{3.7}{0} \mycon{3.7}{-0.5} \myleaf{3.7}{-1}{a^{n - 2} b c} \myhdots{4.6}{0}
            \mymin{5.5}{0} \myleaf{5.5}{-1}{a c^{n - 1}} \myhline{6.5}{0} 
            \mymax{6.6}{0} \mycon{6.6}{-0.5} \myleaf{6.8}{-1}{b c^{n - 1}} \myleaf{7.6}{0}{c^n}
        \end{tikzpicture}$.
\end{itemize}
\subsection{Special reductions} \label{sec:special}
Special condition among shift operators is necessary for some equations to be of polynomial class.  The condition $2a=b+c$ is necessary for the following two examples.
\begin{itemize}
\item[1.]
$F=\begin{tikzpicture}[baseline = -1em, x = 2em, y = 2em]
            \mymin{0}{0} \mymax{1}{0} \myleaf{0}{-1}{a} \myleaf{1}{-1}{b} \myleaf{2}{0}{c}
        \end{tikzpicture}$,\\
$F^n=a^{n - 1}
    \left(
        \begin{tikzpicture}[baseline = -1em, x = 2em, y = 2em]
            \mymin{0}{0} \mymax{1}{0} \myleaf{0}{-1}{a} \myleaf{1}{-1}{b} \myleaf{2}{0}{c}
        \end{tikzpicture}
    \right)$.
\item[2.]
$F=\begin{tikzpicture}[baseline = -1em, x = 2em, y = 2em]
            \mymin{0}{0} \mymin{1}{0} \mycon{0}{-0.5} \mycon{1}{-0.5} \mycon{1.5}{0}
            \myleaf{0}{-1}{a} \myleaf{1}{-1}{b} \myleaf{2}{0}{c}
        \end{tikzpicture}$,
$F^{2n-1}=a^{2n - 2}
    \left(
        \begin{tikzpicture}[baseline = -1em, x = 2em, y = 2em]
            \mymin{0}{0} \mymin{1}{0} \mycon{0}{-0.5} \mycon{1}{-0.5} \mycon{1.5}{0}
            \myleaf{0}{-1}{a} \myleaf{1}{-1}{b} \myleaf{2}{0}{c}
        \end{tikzpicture}
    \right)$,
$F^{2n}=a^{2n-2}
\left(
        \begin{tikzpicture}[baseline = -2em, x = 2em, y = 2em]
            \mymax{0}{0} \myleaf{0}{-1}{a^2} \mymin{1}{0} \mymax{1}{-1} \myleaf{1}{-2}{ab} \myhline{2}{0} \myleaf{2}{-1}{b^2} \mymax{3}{0} \myleaf{3}{-1}{ac} \mymin{4}{0} \myleaf{4}{-1}{ab} \myleaf{5}{0}{c^2}
        \end{tikzpicture}
    \right)
$.
\end{itemize}

We can extend the reduction in the case 3 of subsection \ref{sec:P1}. Replace the term $c$ by $c\wedge\overline{d}$ of $F$ of the case, then we obtain the following relation for $F^2$.
\begin{equation*}
    \begin{aligned} 
        &\left(
            \begin{tikzpicture}[baseline = -1em, x = 2em, y = 2em]
                \mymin{0}{0} \myleaf{0}{-1}{a} \mymax{1}{0} \mycon{1}{-0.5} \myleaf{1}{-1}{b} \mymin{2}{0} \myleaf{2}{-1}{c} \mycon{2.5}{0} \myleaf{3}{0}{d}
            \end{tikzpicture}
        \right)^2 \\=
        &\left(
            \begin{tikzpicture}[baseline = -1em, x = 2em, y = 2em]
                \mymin{0}{0} \myleaf{0}{-1}{a} \mymax{1}{0} \mycon{1}{-0.5} \myleaf{1}{-1}{b} \myleaf{2}{0}{c}
            \end{tikzpicture}
        \right)
        \left(
            \begin{tikzpicture}[baseline = -1em, x = 2em, y = 2em]
                \mymin{0}{0} \myleaf{0}{-1}{a} \mymax{1}{0} \mycon{1}{-0.5} \myleaf{1}{-1}{b} \mymin{2}{0} \myleaf{2}{-1}{c} \mycon{2.5}{0} \myleaf{3}{0}{d}
            \end{tikzpicture}
        \right).
    \end{aligned}
    \end{equation*}
Using this relation repeatedly, we have
\begin{equation} \label{special}
    \begin{aligned} 
        &\left(
            \begin{tikzpicture}[baseline = -1em, x = 2em, y = 2em]
                \mymin{0}{0} \myleaf{0}{-1}{a} \mymax{1}{0} \mycon{1}{-0.5} \myleaf{1}{-1}{b} \mymin{2}{0} \myleaf{2}{-1}{c} \mycon{2.5}{0} \myleaf{3}{0}{d}
            \end{tikzpicture}
        \right)^n \\=
        &\underbrace{\left(
            \begin{tikzpicture}[baseline = -1em, x = 2em, y = 2em]
                \mymin{0}{0} \myleaf{0}{-1}{a} \mymax{1}{0} \mycon{1}{-0.5} \myleaf{1}{-1}{b} \myleaf{2}{0}{c}
            \end{tikzpicture}
        \right)^{n-1}}_{\text{class $P_1$}}
        \left(
            \begin{tikzpicture}[baseline = -1em, x = 2em, y = 2em]
                \mymin{0}{0} \myleaf{0}{-1}{a} \mymax{1}{0} \mycon{1}{-0.5} \myleaf{1}{-1}{b} \mymin{2}{0} \myleaf{2}{-1}{c} \mycon{2.5}{0} \myleaf{3}{0}{d}
            \end{tikzpicture}
        \right).
    \end{aligned}
\end{equation}
Since $(a \wedge (\overline{b} \vee c))^{n-1}$ is in class $P_1$ as shown in the case 3 of subsection \ref{sec:P1}, $(a \wedge (\overline{b} \vee (c \wedge \overline{d})))^n$ is also in class $P_1$ by \eqref{special}. If we extend this mechanism, we obtain the following reduction relations.  Note that the complexity class depends on $X$ and it is equal to the class $P$ of $(n-1)$-th power in the right-hand side.  For example, if $X=c$, both classes in the case 3 and 4 become $P_1$.
\begin{itemize}
    \item [3.]
    \begin{equation*}
        \begin{aligned}
            &\left(                    
                \begin{tikzpicture}[baseline = -1em, x = 2em, y = 2em]
                    \mymin{0}{0} \myleaf{0}{-1}{a_1} \myhdots{1}{0} \mymin{2}{0} \myleaf{2}{-1}{a_m} 
                    \mymin{3}{0} \mycon{3}{-0.5} \myleaf{3}{-1}{b_1} \myhdots{4}{0} \mymin{5}{0} \mycon{5}{-0.5} \myleaf{5}{-1}{b_n} \myleaf{6}{0}{X}
                \end{tikzpicture}
            \right)^n \\ =
            &\underbrace{\left(                    
                \begin{tikzpicture}[baseline = -1em, x = 2em, y = 2em]
                    \mymin{0}{0} \myleaf{0}{-1}{a_1} \myhdots{1}{0} \mymin{2}{0} \myleaf{2}{-1}{a_m} \myleaf{3}{0}{X}
                \end{tikzpicture}
            \right)^{n-1}}_{\text{class $P$}} \\
            &\quad\left(
                \begin{tikzpicture}[baseline = -1em, x = 2em, y = 2em]
                    \mymin{0}{0} \myleaf{0}{-1}{a_1} \myhdots{1}{0} \mymin{2}{0} \myleaf{2}{-1}{a_m} 
                    \mymin{3}{0} \mycon{3}{-0.5} \myleaf{3}{-1}{b_1} \myhdots{4}{0} \mymin{5}{0} \mycon{5}{-0.5} \myleaf{5}{-1}{b_n} \myleaf{6}{0}{X}
                \end{tikzpicture}
            \right)       
        \end{aligned}
    \end{equation*}
    \item [4.]
    \begin{equation*}
        \begin{aligned}
            &\left(
                \begin{tikzpicture}[baseline = -1em, x = 2em, y = 2em]
                    \mymin{0}{0} \myleaf{0}{-1}{a} \mymax{1}{0} \mycon{1}{-0.5} \myleaf{1}{-1}{b} \mymin{2}{0} \mycon{2}{-0.5} \myleaf{2}{-1}{c_1} \myhdots{3}{0} \mymin{4}{0} \mycon{4}{-0.5} \myleaf{4}{-1}{c_2} \myleaf{5}{0}{X}
                \end{tikzpicture}
            \right)^n \\=
            &\underbrace{\left(
                \begin{tikzpicture}[baseline = -1em, x = 2em, y = 2em]
                    \mymin{0}{0} \myleaf{0}{-1}{a} \mymax{1}{0} \mycon{1}{-0.5} \myleaf{1}{-1}{b} \myleaf{2}{0}{X}
                \end{tikzpicture}
            \right)^{n-1}}_{\text{class $P$}} \\
            &\quad\left(
                \begin{tikzpicture}[baseline = -1em, x = 2em, y = 2em]
                    \mymin{0}{0} \myleaf{0}{-1}{a} \mymax{1}{0} \mycon{1}{-0.5} \myleaf{1}{-1}{b} \mymin{2}{0} \mycon{2}{-0.5} \myleaf{2}{-1}{c_1} \myhdots{3}{0} \mymin{4}{0} \mycon{4}{-0.5} \myleaf{4}{-1}{c_2} \myleaf{5}{0}{X}
                \end{tikzpicture}
            \right)
        \end{aligned}
    \end{equation*}
\end{itemize}
\section{Concluding remarks}
We have proposed a new type of lattice equations of polynomial class using the novel type of expressions for operators such as shift operators and binary trees.  Finally, we give some remarks as follows.
 
Some of ECA's are included as a special case of operators and state variables.  If we assume $A\vee B=\max(A,B)$, $A\wedge B=\min(A,B)$, $\overline{A}=1-A$, then the state variable is closed in the binary value 0 and 1.  Moreover, if we assume a limit for shift operators from $-1$ to 1, we can obtain ECA's from lattice equations.  For example, ECA12 is obtained if $(a,b)=(0,-1)$ for the case 2 of subsection \ref{sec:P0}. ECA138, 140 and 162 are obtained if $(a,b,c)=(1,-1,0)$, $(0,-1,1)$ and $(1,0,-1)$ respectively for the case 3 of subsection \ref{sec:P1}.
  
All equations of polynomial class shown in the reference \cite{ikegami} follow the form $u_i^{n+1}=f(u_{i-1}^n,u_i^n,u_{i+1}^n)$.  Those equations can be analyzed by our approach and the same results can be derived.  However, shift operators are arbitrary for the above results in this letter.  Therefore, the results reported in this letter cover and extend those in the reference \cite{ikegami}.

There is another type of equations and their exact solutions regarding the max operator.  The semi-field defined by basic operations $\max$ (or $\min$), $+$ and $-$ is called `max (min)-plus semi-field'\cite{heidergott}.  Our approach cannot be applied to max-plus equations and their solutions since the set of basic operations are different from lattice operations. However, some solutions to specific max-plus equations are quite familiar with lattice operations.  For example, those to max-plus extensions of particle cellular automata in ref.~\cite{particle} are described by $\max(A,B)$, $\min(A,B)$ and $1-A$ which can be lattice operations on $\mathbf{R}$ though rational constants are also included.  Such constants are not treated in reductions proposed of this letter.  However, some extension to our reductions may be applied to such constants since comparison of magnitude of numbers can be introduced into our approach.
\references

\end{document}